\documentclass[prl,aps,epsf,showpacs,twocolumn]{revtex4}
\usepackage{times}
\usepackage{graphicx}
\usepackage{float}
\usepackage{latexsym,amsmath,amssymb,bm,euscript}
\usepackage{color}
\usepackage{subfigure}
\usepackage{epstopdf}
\usepackage[colorlinks=true,linkcolor=blue,citecolor=blue]{hyperref}
\usepackage{hyperref}

\begin{document}

\title{Scaling behavior of the insulator to  plateau transition in topological band model}
\author{Jerimie Priest, S. P. Lim and D. N. Sheng} 
\affiliation {Department of Physics and Astronomy, California State University, Northridge, California 91330, USA}

\begin{abstract} 
Scaling behavior of the quantum phase transition from an insulator to a quantum Hall
plateau state  has often been examined within systems realizing Landau levels.
We study the topological transition in  energy band model with nonzero Chern number,
which has  same topological property as a Landau level.
We find that  topological  band generally realizes the same universality class as the integer quantum Hall system
under uniform magnetic flux for strong enough disorder scattering. 
Furthermore,   the symmetry of the transition characterized by  relations:
$\sigma_{xy}(E)=1-\sigma_{xy}(-E)$ for  Hall conductance  and $\sigma_{T}(E)=\sigma_{T}(-E)$
for  longitudinal Thouless conductance is  observed near the transition region.
We also establish that  finite temperature dependence of  Hall conductance is  determined
by  inelastic scattering relaxation time, while the  localization exponent $\nu$ remains unchanged
by such scattering. 
\end{abstract}

\maketitle

{\bf Introduction--} Anderson localization  theory\cite{anderson1, pal_review, mac} predicts that  noninteracting electrons are generally  
localized in two-dimensional (2D) disordered systems at zero temperature limit without a magnetic field. 
In contrast to the physics  of  localization for such systems belonging to the orthogonal class,
extended (delocalized) states\cite{laughlin, pruisken,huo}  exist in systems with a magnetic field or spin-orbit coupling\cite{spinorbit}, which are capable
of  conducting  electric current going through whole samples.
The topological characterization\cite{huo,thouless_top,koh, aro, dns_so} plays a central role in understanding 
delocalization of these systems, where topological nontrivial states characterized by  nonzero
Chern numbers must exist associated with  delocalization  in these  2D systems.
The integer quantum Hall effect (IQHE) discovered for electron systems  under magnetic field demonstrates
universal scaling behavior\cite{exp} in accordance with a single delocalized quantum critical point ($E_c$)\cite{pruisken,huo,ando,huk,chalker} between
a trivial insulator  and a quantized Hall plateau  state or between two adjacent plateau states,  
while  localization length diverges  as a power law $\xi\sim |E_f- E_c|^{-\nu}$ near the quantum critical point\cite{ando, huk, chalker,con, lattice}.
Physical quantities like conductances follow the one parameter scaling law  as a function of the scaling variable
$(L/\xi)^{1/\nu} \propto L^{1/\nu}\Delta E$ with $\Delta E=E_f-E_c$ and $L$ being the system length. 
However, experiments have been done at finite temperature where  near thermodynamic sample size is being cut-off by
a finite length scale $L_{in}\propto T^{-p/2}$ representing the dephasing effect\cite{thouless1, anderson2, review} with $p$ known as the  inelastic
scattering exponent.  Thus the scaling parameter for finite temperature systems 
appear to be $T^{-\kappa}\Delta E$ with $\kappa=p/2\nu$\cite{exp}.
While extensive numerical studies establish a robust universal scaling dependence on the system length
with the scaling exponent $\nu\sim 2.40-2.60$ (more recent studies suggest slightly bigger
value\cite{new} than earlier results\cite{huk,chalker, con,lattice,dns1, review}).
The same exponent $\nu$ has also been established\cite{lattice} for lattice models  with  uniform magnetic flux.
On the other hand,   there are no well established results for understanding the temperature 
scaling law\cite{exp, review} in a microscopic model incorporating disorder and inelastic scattering effects. 
For noninteracting system,  recent numerical studies based on noncommutative Kubo formula support
that with the input of $1/\tau_{in}\propto kT$, one obtains  a $\kappa=1/2\nu\sim 0.2$\cite{emil1}.
However,  the physical original for the finite temperature scaling behavior remains not well understood. 
Very interestingly,  experimentally observed decay exponent $\kappa \simeq 0.43$ is
likely to be universal  suggesting the scenario that interaction must play very important role in these systems.    
Since the scaling behaviors for interacting and disordered systems are much harder to   be settled\cite{review},  
it is highly interesting to examine 
precisely how the inelastic scattering controls the temperature scaling of  transport in noninteracting systems, 
which may  provide a systematic understanding to  experimental observations.

\begin{figure}[b]
  \includegraphics[width=2.5in]{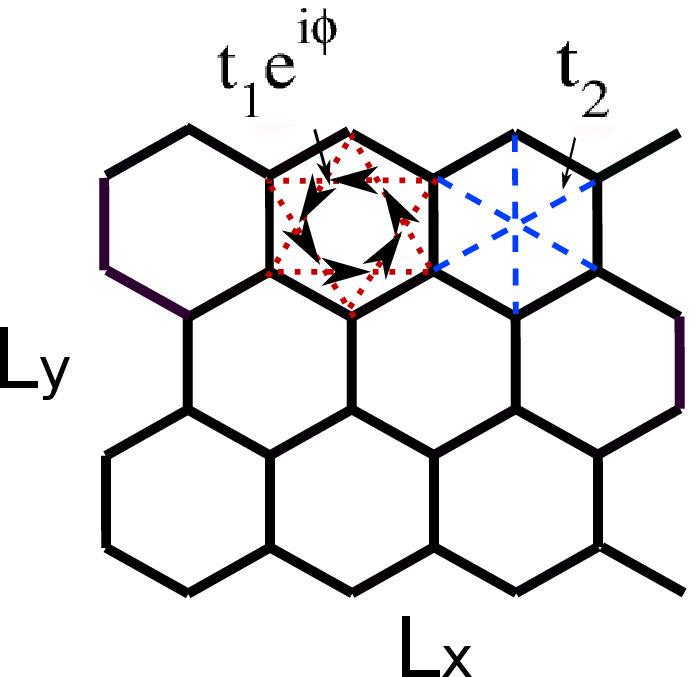}
 \includegraphics[width=2.5in]{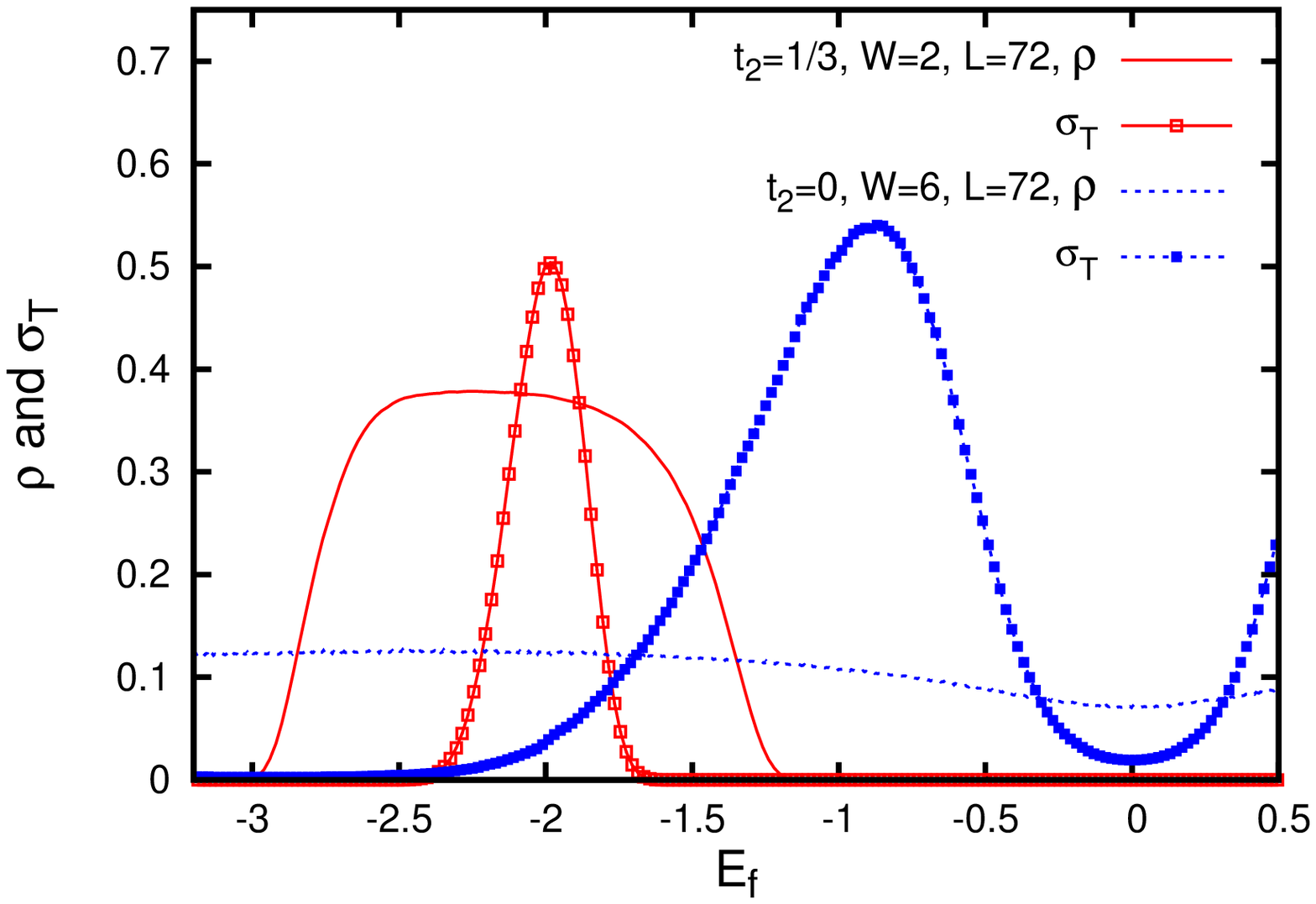}
\hspace{-0.0in}
\vspace{-0.92in}
\caption{(Color online) (a) 
The Haldane model on the honeycomb lattice for $L_x=8$ and
$L_y=4$. The arrow directions represent the 
positive hopping phase $\phi$ in  next NN hoppings (with magnitude $t_1$).
The third neighbor hoppings ($t_2$) are represented by  blue dashed lines.
(b) The density of states $\rho$ and Thouless conductance $\sigma_{T}$
for two cases  representing different strengths
of disorder and broadening of the topological band
 (with $t_2$=1/3, $W=2$ and   $t_2=0$,  $W=6$ respectively). 
The relatively sharp feature in $\sigma_{T}$ indicates Anderson
localization away from the peak of the conductance.
 }
\label{density}
\end{figure}
                                 
On a different forefront,   topological band models first proposed by Haldane\cite{haldane} have attracted
a lot of recent attention since  energy bands of such systems can be tuned to  mimic  uniform magnetic field,  and  host nontrivial
fractionalized topological states\cite{flatband} despite the net zero flux in the system.
Such models also realize the $Z_2$ topological quantum spin Hall effect\cite{kane, bernevig, dns_qshe}   once we include  spin degrees of freedom,  which have also been  extensively studied 
for their interesting delocalization properties\cite{lattice1, lattice2, lattice3, lattice4,lattice5}. 
There are still open  questions regarding  universal properties of the insulator to quantum Hall state transition in such systems. 
It is unclear if such systems will realize the same universality class as the
system under uniform magnetic flux  since 
 a region of extended states may   exist based on the  argument of
 possible delocalized critical states in random flux problem\cite{random}. 
One can also ask what is characteristic length scale to reach the universal scaling law for such systems if they do satisfy it in  thermodynamic limit.
 In particular,  we are interested in exploring universal  behaviors of the transition for both system length dependence at zero temperature  and finite temperature
scaling, which have not been studied  simultaneously in microscopic model
simulations so far.

In this paper, we present numerical study 
based on  calculations of  longitudinal Thouless conductance\cite{thouless2}
$\sigma_T$  and Hall conductance $\sigma_{xy}$ using
the exact diagonalization method for  topological band model.
 Our results show that a direct insulator to plateau  transition
indeed belongs to the same universality class as the conventional IQHE system when disorder scattering is strong.
In particular, the scaling exponent $\nu$ is universal and  scaling curves for
both $\sigma_{T}$ and $\sigma_{xy}$ respect the particle-hole symmetry for large enough  system length ($L \sim 120$), 
which continuously adjust themselves towards recovering the symmetry with the increase of $L$. 
The temperature scaling behavior is fully determined by the inelastic scattering relaxation time
$1/\tau_{in}$ at low temperature limit, which results in a universal scaling law
 $d\sigma_{xy}/dE_f\simeq (1/\tau_{in})^{-1/2\nu}$ for finite temperature conductance.
While this is consistent with the scaling theory\cite{pruisken,emil1},  to our best knowledge,  it is for the first 
time being revealed
based on numerical model simulations using  conventional Kubo formula. 
Our results are consistent with the experimental finding that the electron-electron interaction
only comes into play as a temperature dependent relaxation time ($1/\tau_{in}\propto (kT)^2$),
which does not change the localization length exponent $\nu$. 

{\bf Lattice model and method---}We study the Haldane model~\cite{haldane} on the honeycomb  lattice
in the following tight-binding form:
\begin{eqnarray*} 
H=(-t\sum_{<ij> }c_i^+c_j+t_1\sum_{<<ij>> } e^{i \phi_{ij}}c_i^+c_j\\
+t_2\sum_{<<<ij>>> } c_i^+c_j 
 + H.c.) +\sum _i w_i c^+_i c_i 
\end{eqnarray*} 
where $\langle\dots\rangle$, $\langle\langle\dots\rangle\rangle$ and
$\langle\langle\langle\dots\rangle\rangle\rangle$ denote the nearest-neighbor (NN), the
next NN and the third neighbor of pairs of sites.
The  $c_i^+$ is a  fermionic creation operator 
and $w_i$ is  a disorder  potential uniformly distributed
between $(-W/2, W/2)$.
We set  $t=1$, $t_1=0.40$,
and vary $t_2$ as a parameter to tune the band width of the 
energy spectrum, which can realize a flatband\cite{ywang} at $t_2=1/3$ and wider band at $t_2=0$. 
We study finite size system with $N=L_xL_y$ sites, where $L_x$
represents the number of sites along each zigzag chain and
$L_y$ is the number of chains as illustrated in Fig. \ref{density}(a).
In our simulation, we  will set $L_x=L_y=L$.

\begin{figure}[b]
 \includegraphics[width=2.5in]{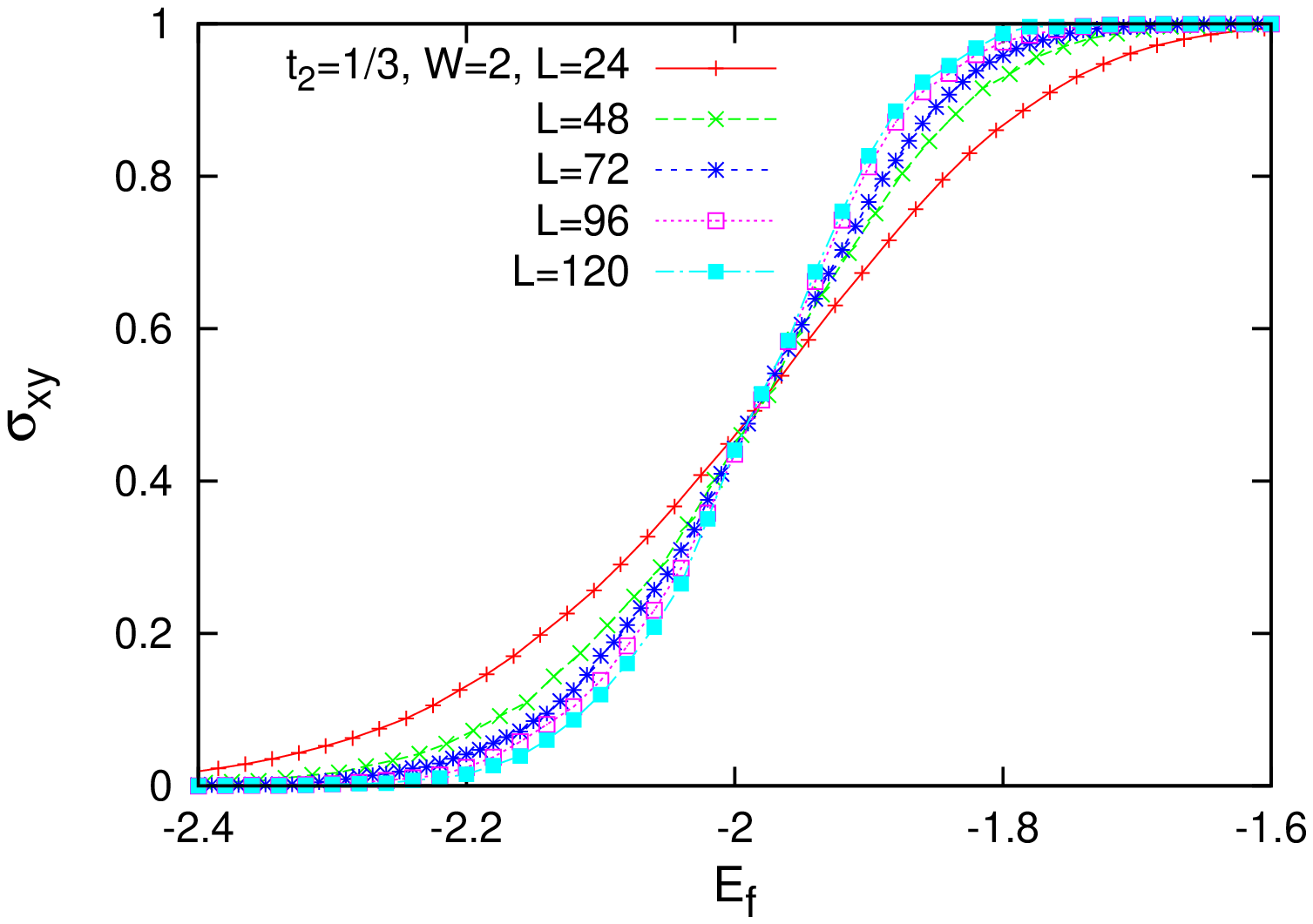}
  \includegraphics[width=2.5in]{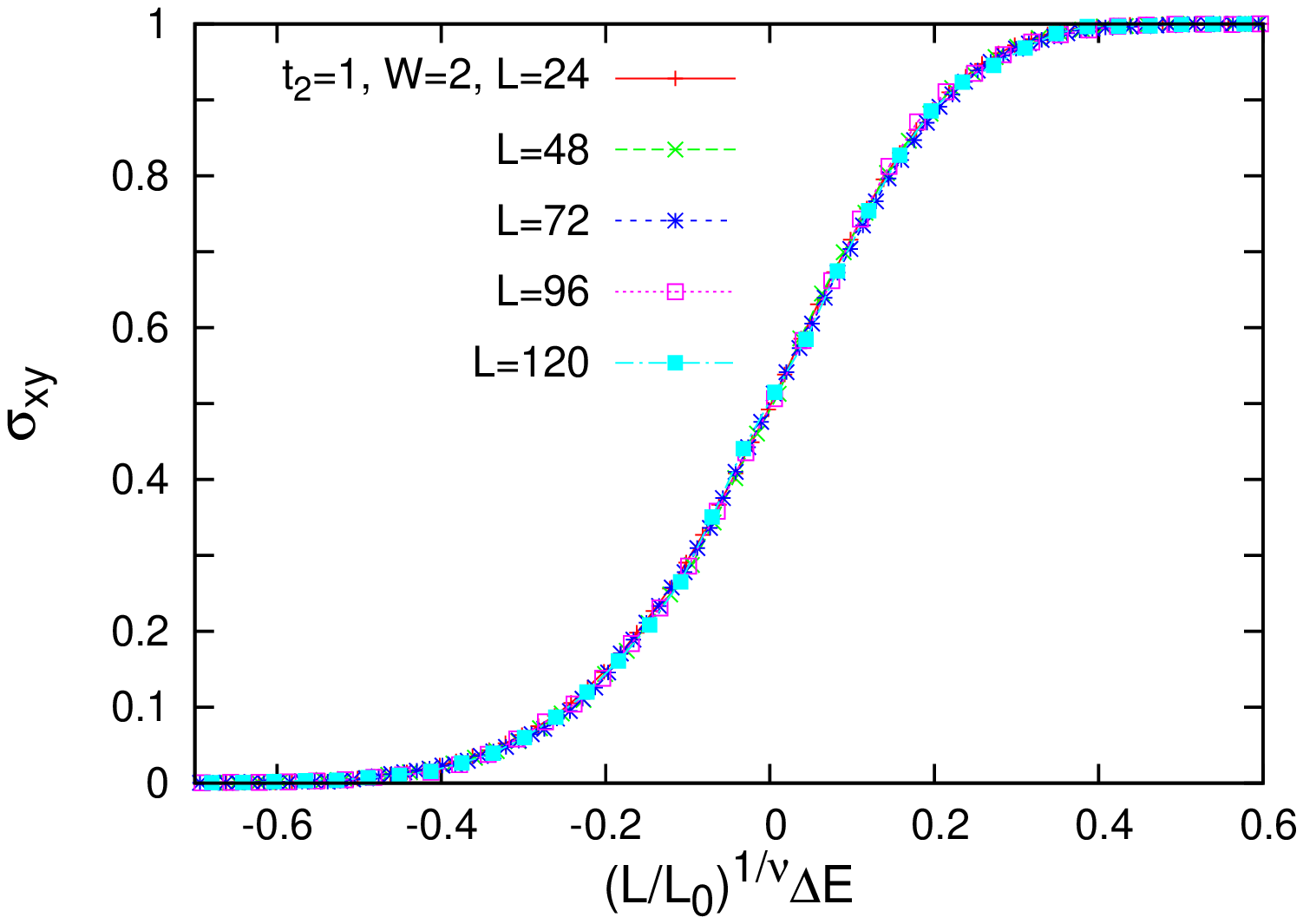}
\hspace{-0.0in}
\vspace{-2.1in}
\caption{(Color online) (a) The evolution of Hall conductance $\sigma_{xy}$
with Fermi energy $E_f$ for different system sizes from
$L=24$ to $120$ for the model with $t_2=1/3$ and $W=2$. All data points cross at a critical
energy $E_c$ and the transition  to  $\sigma_{xy}=e^2/h$ Hall plateau
 is becoming sharper with the increase
of the $L$. (b) Using
the scaling variable $(L/L_0)^{1/\nu}\Delta E$ ($\Delta E=E_f-E_c$) to rescale
all curves. All  data for different $L$ collapse  into one smooth curve  consistent
with  one parameter scaling law for the IQHE. We find  best fitting results
using $\nu=2.5$. We choose the  parameter $L_0=24$.
 }
\label{hall1}
\end{figure}

\begin{figure}[b]
  \includegraphics[width=2.2in]{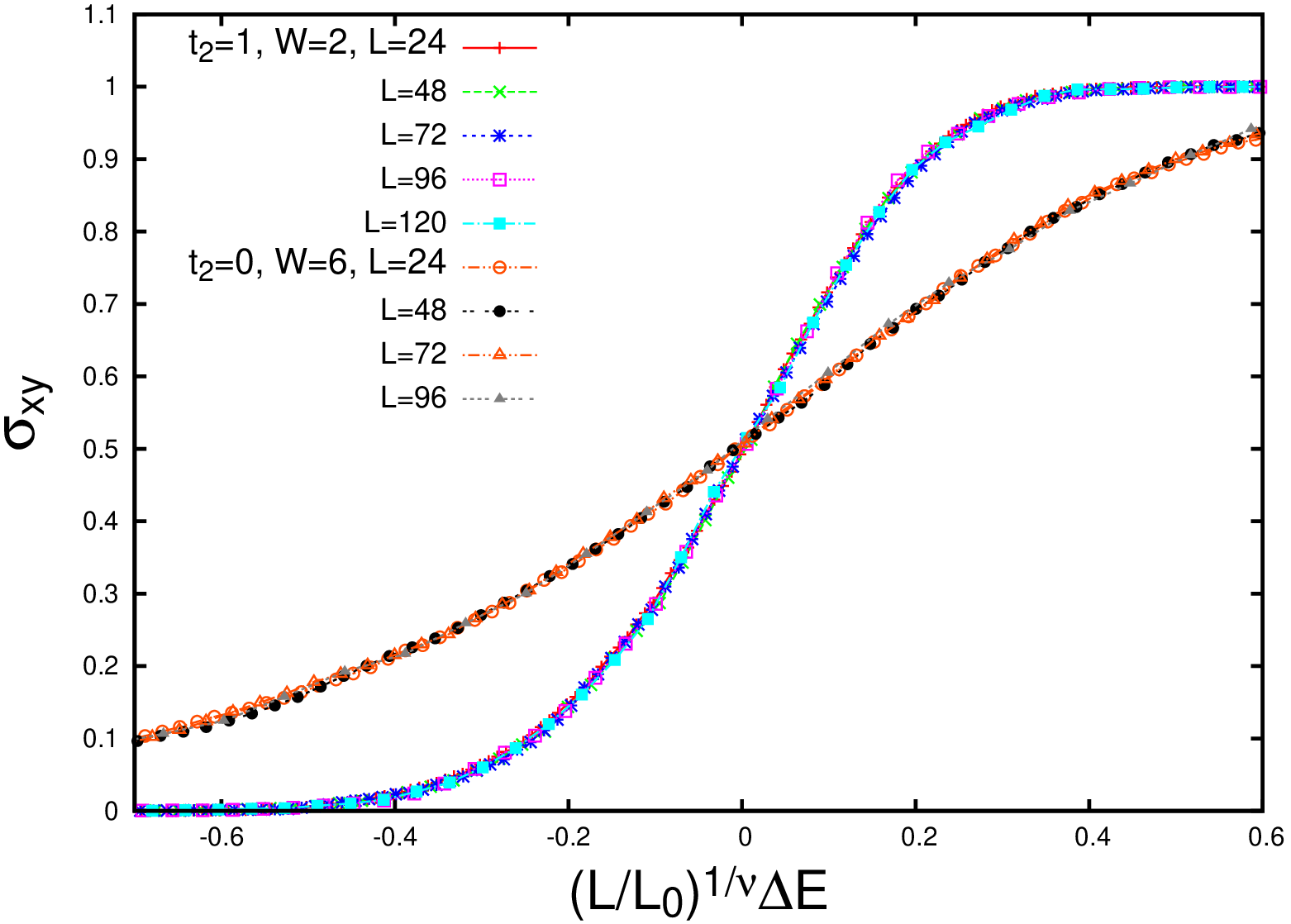}
  \includegraphics[width=2.17in]{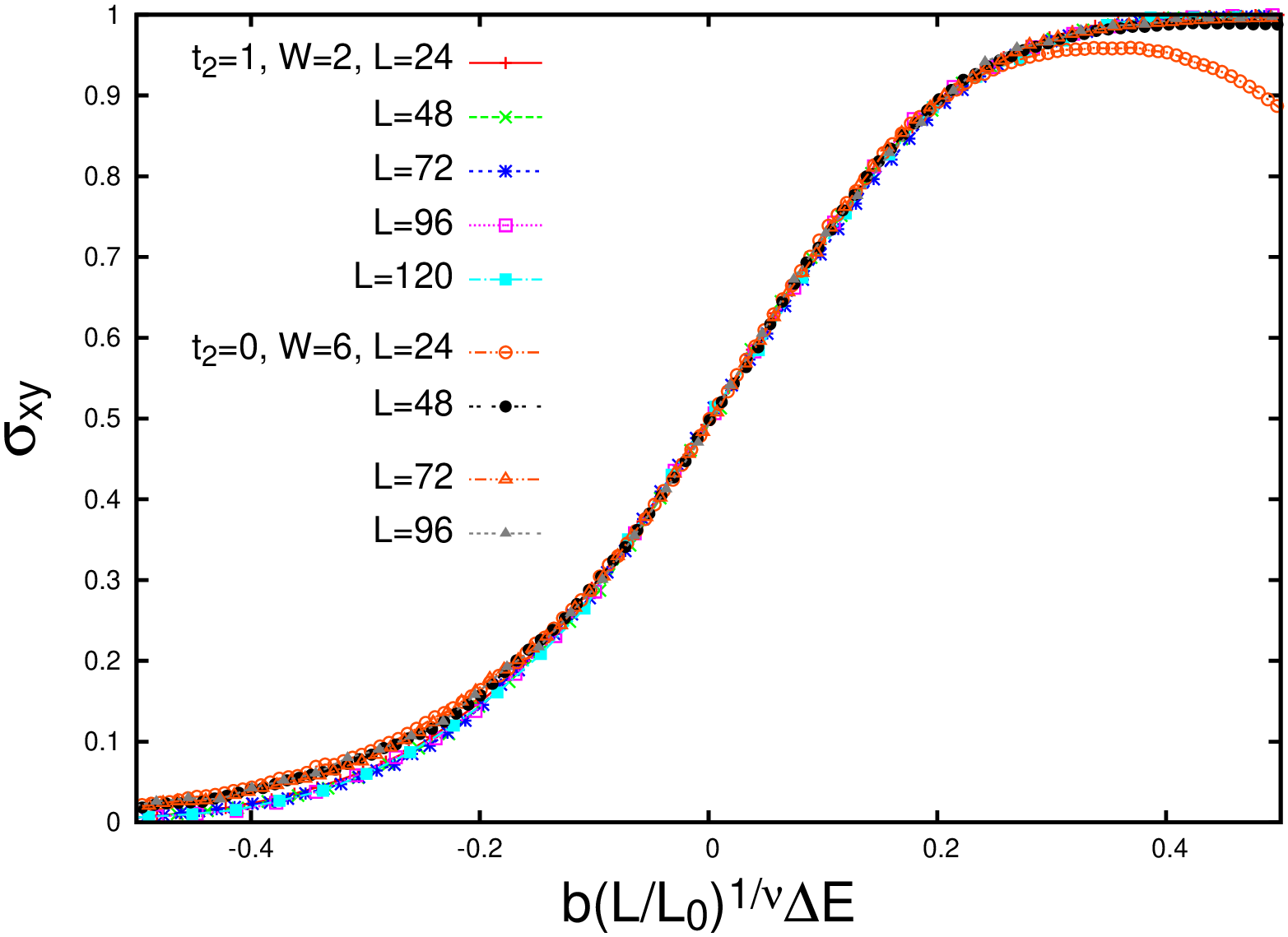}
\hspace{-0.0in}
\vspace{-1.7in}
\caption{(Color online) (a) $\sigma_{xy}$ for
 both $W=2$ and $W=6$  
are  shown to follow similar scaling curves
 with the scaling variable $(L/L_0)^{\nu}\Delta E$.
The same $\nu=2.5$ is obtained for the best collapsing effect  of all  data.
We find that the transition in the  wider band case ($W=6$) is much
slower.
(b)  The two scaling curves can be rescaled together by
adjusting the scaling  variable by a constant $b$ to the $W=6$ curve.
We obtain $b=0.40 $,  indicating the length scale  in the latter is
relatively bigger as $L_0/b^{\nu}$. 
 }
\label{hall2}
\end{figure}

\begin{figure}[b]
 \includegraphics[width=2.2in]{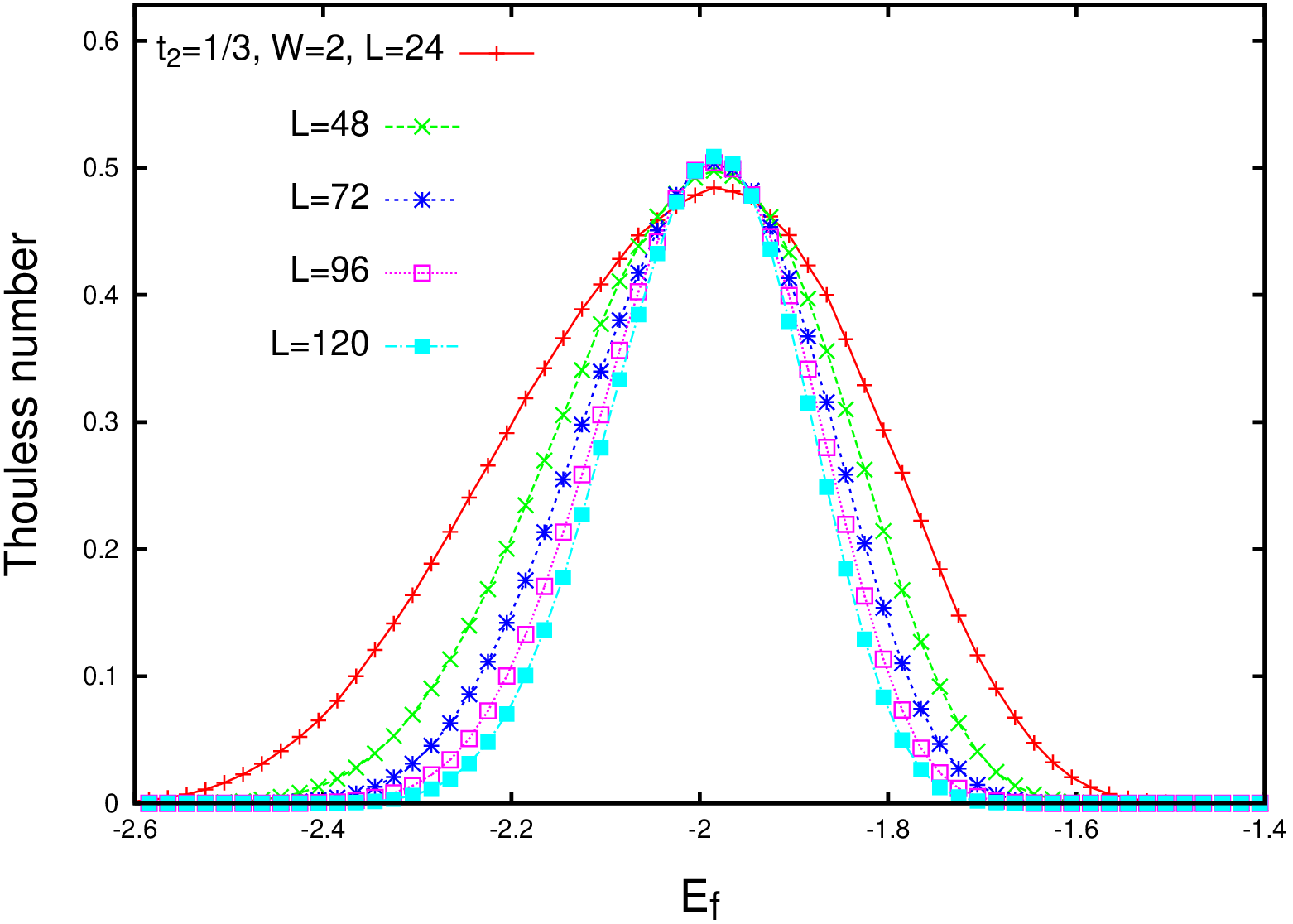}
  \includegraphics[width=2.2in]{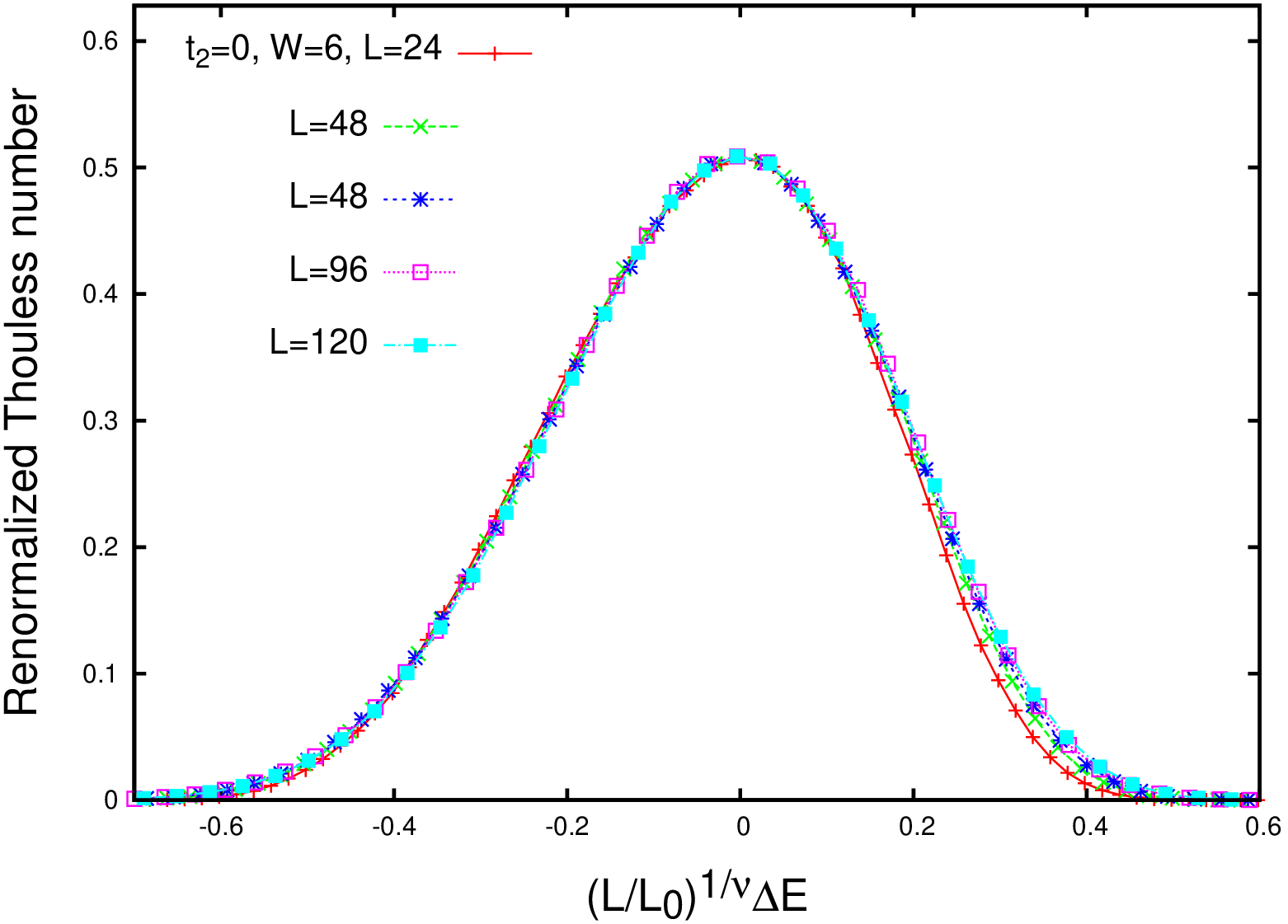}
\hspace{-0.0in}
\vspace{-1.7in}
\caption{(Color online) (a) The Thouless conductance $\sigma_{T}$ data
for system length $L=24$ to $L=120$ for $t_2=1/3$ and $W=2$. The peak appears 
at the same critical energy $E_c$ identified as the crossing point of
the $\sigma_{xy}$ for different $L$.  
The width of each curve shrinks as $L$ increases.  The $\sigma_{T}$
at the peak grows slowly with $L$, which scales to a constant
at thermodynamic limit.  (b) The renormalized $\sigma_{xx}$
is shown to follow the one parameter scaling law with the same $\nu=2.5$.
All  data from larger sizes (except $L=24$ case) collapse nicely into one curve.
 }
\label{thouless1}
\end{figure}

\begin{figure}[b]
 \includegraphics[width=2.2in]{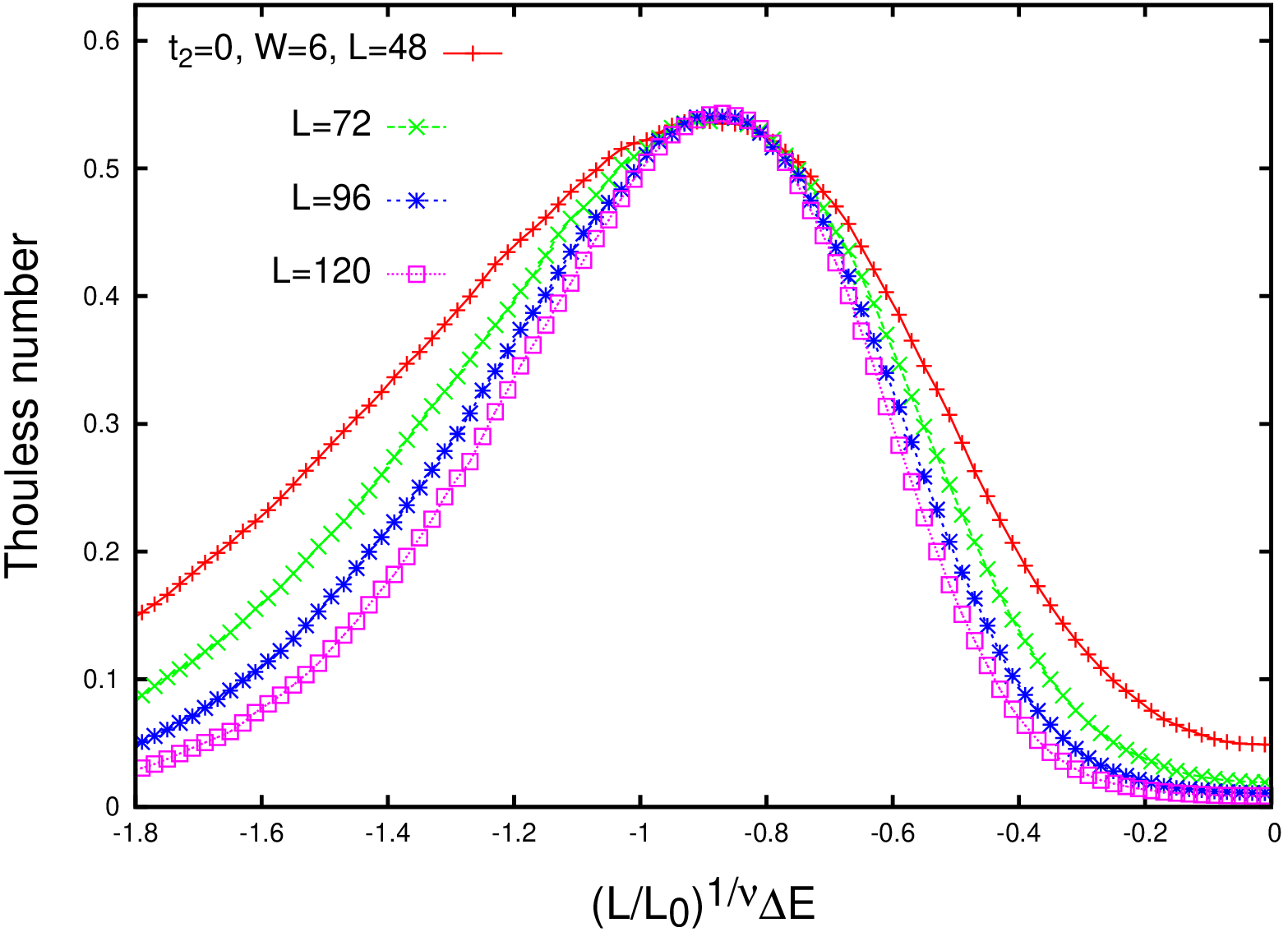}
  \includegraphics[width=2.2in]{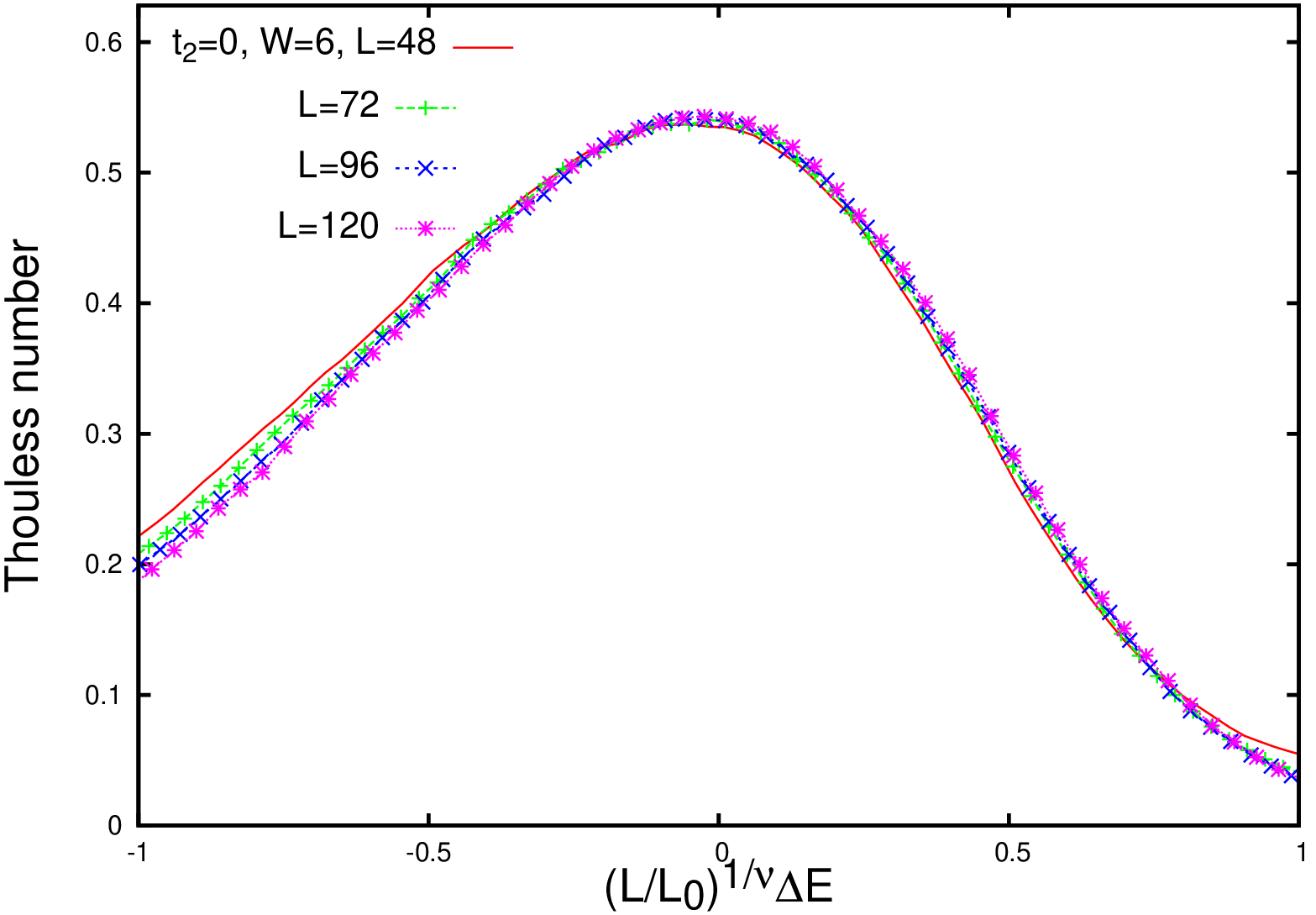}
\hspace{-0.0in}
\vspace{-1.7in}
\caption{(Color online) (a) The Thouless conductance $\sigma_{T}$
for system sizes $L=48$ to $L=120$ for $t_2=0$ and $W=6$. 
(b) There are stronger finite size effect for this system due
to much larger length scale. We find all  data can only be approximated
fitting into one curve.   Interestingly,  the larger sizes 
results ($L=120$) are systematically deviating from the curves for
smaller $L$,  as $\sigma_{T}$  drops faster at $\Delta E<0$ side and slower at the other side to adjust towards more symmetric curve.
 }
\label{thouless2}
\end{figure}

\begin{figure}[b]
\includegraphics[width=2.5in]{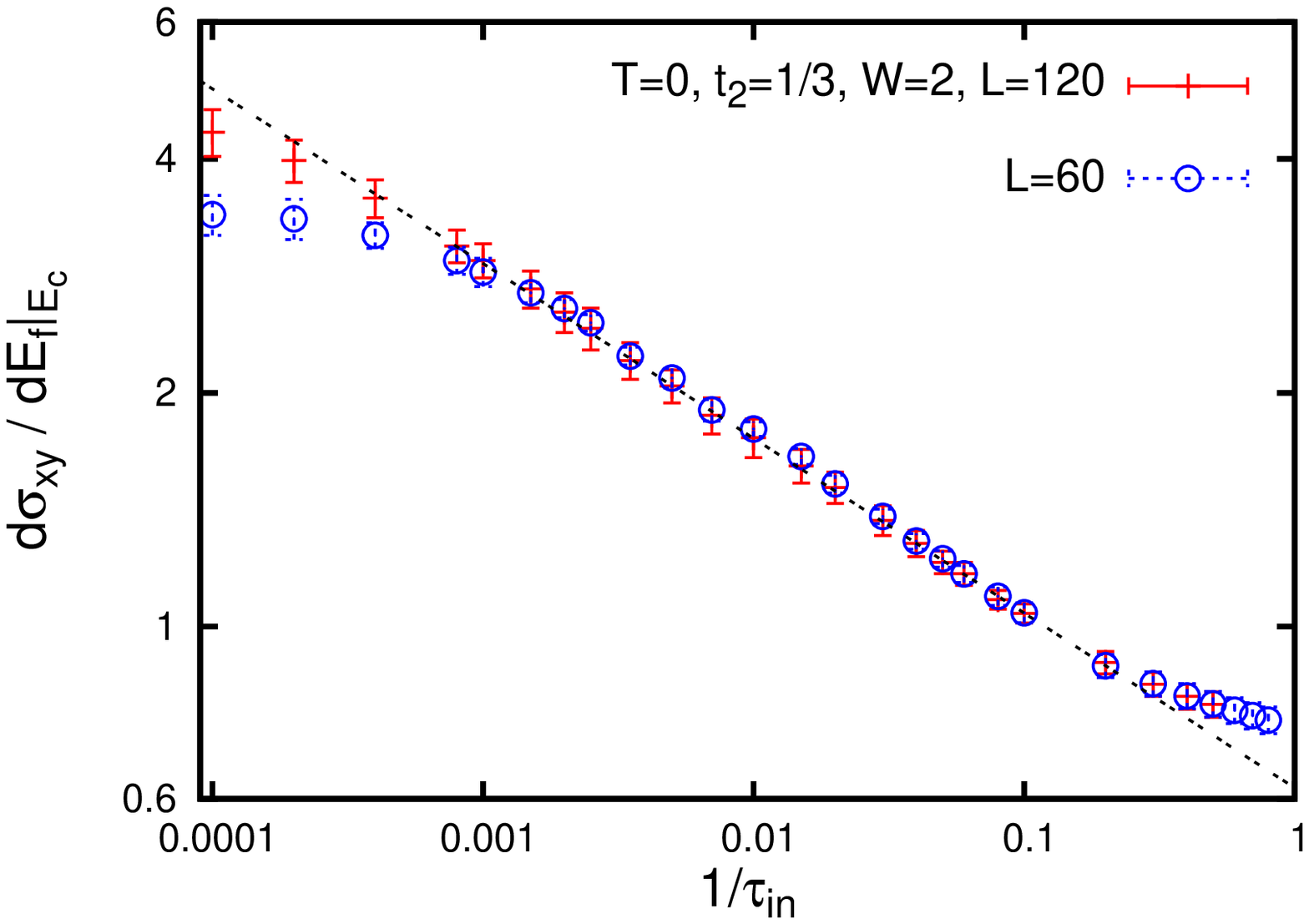}
\includegraphics[width=2.5in]{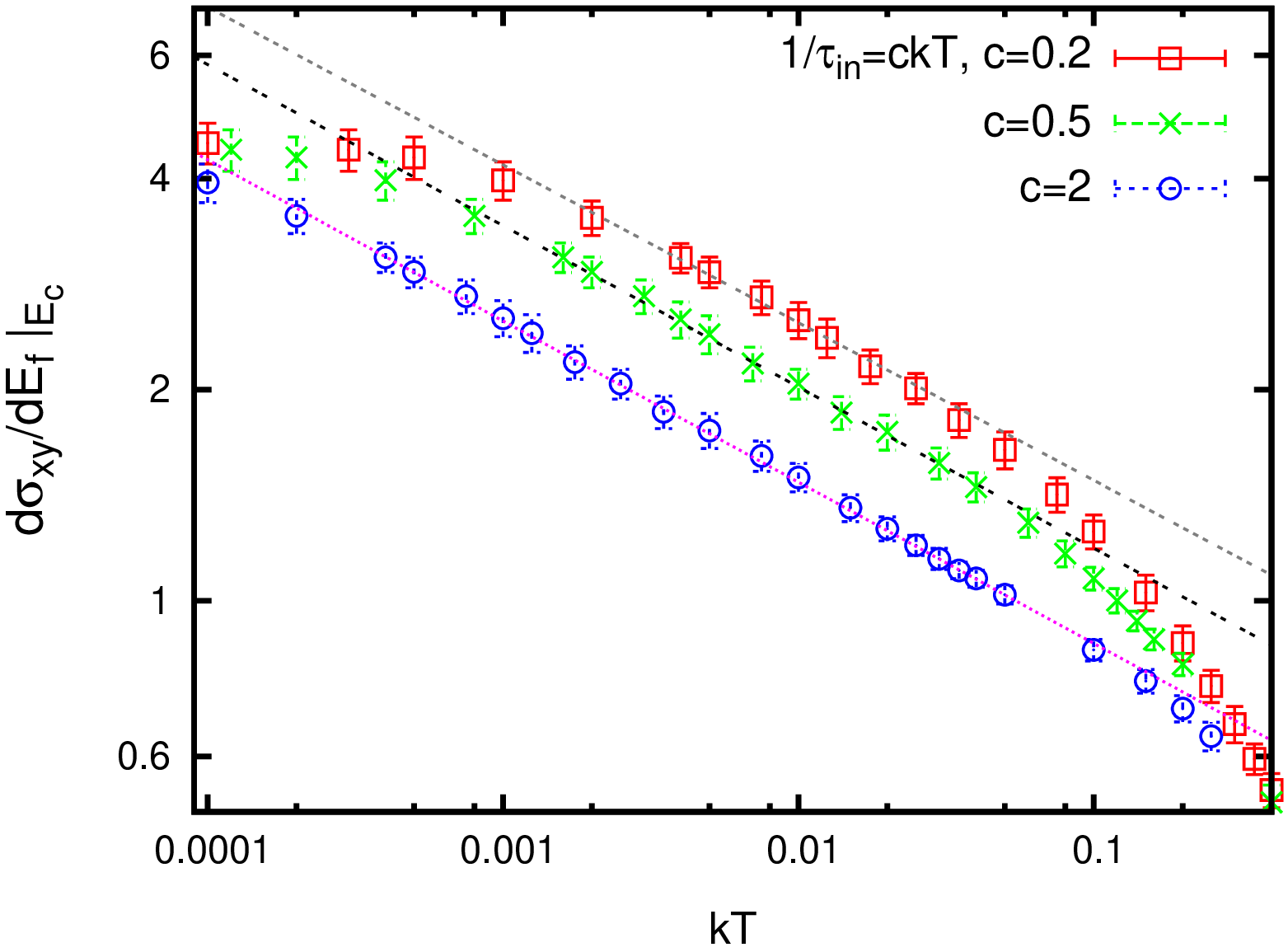}

\hspace{-0.0in}
\vspace{-2.3in}
\caption{(Color online) (a)
The scaling behavior of the $(d\sigma_{xy}/dE_f)|{E_c}$ as a function of the inverse of the relaxation time
for two systems with $L=120$ and $60$ at $T=0$.  (b) The $(d\sigma_{xy}/dE_f)|{E_c}$ for
finite $T$ $\sigma_{xy}$ as a function of $kT$.
The relaxation time has been chosen as $1/\tau_{in}=ckT$, where the contribution
from the Fermi-Dirac distribution sets in at higher $T$.
 }
\label{eta}
\end{figure}

\begin{figure}[b]
\includegraphics[width=2.3in]{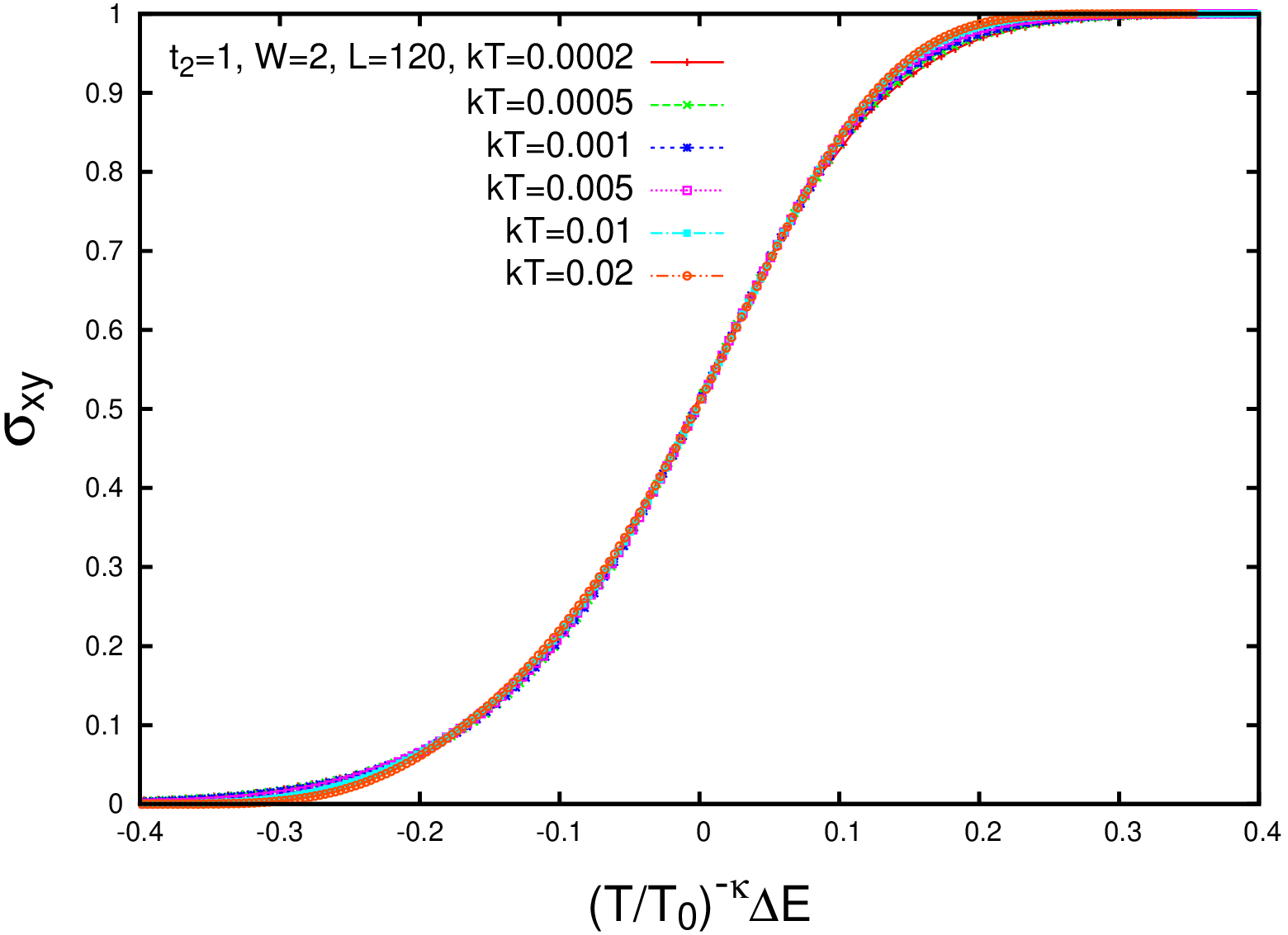}
\hspace{-0.0in}
\vspace{-1.3in}
\caption{(Color online)  
The finite temperature $\sigma_{xy}$ for a wide range of $kT$ between $0.0002$ to $0.02$ (in units
of $t$) has been shown as the function of the scaling variable $(T/T_0)^{-\kappa}\Delta E$ ($kT_0=0.0002$),
where all  data collapse onto one curve, with best fitting exponent $\kappa=0.22\simeq 1/2\nu$.
 }
\label{eta-T}
\end{figure}

To  examine  critical behavior of the system
for different energy  bands with different band broadening,
we first show  two examples of densities of states $\rho$ in Fig. \ref{density}(b).
We obtain  longitudinal conductance from  Thouless number calculations\cite{thouless2,mac1}
as  $\sigma_{T}= 2\pi E_T/ \Delta$, where $E_T$ is the geometry
average of the  energy difference for eigenstates between periodic and antiperiodic boundary
conditions and $\Delta$ is  average energy level spacing.
For $t_2=1/3$, which realizes a near flatband\cite{ywang} without disorder, 
we can see that two bands separated by a gap 
as shown in Fig. \ref{density}(a)
(the spectrum is symmetric about $E=0$ and we only show one band), each carrying a nonzero but opposite 
Chern number $C=\pm 1$\cite{haldane,ywang}, respectively. The $\rho$ is still well 
broadened by random disorder at $W=2$.  For another case
with  $t_2=0$ and $W=6$,  the two bands are mixed
together  and  $\rho$ is near constant in  whole energy range due to  strong disorder
effect.  We note that  longitudinal conductance peaks are very sharp comparing to the profile of the densities of states, indicating that most of states
are localized according to Anderson localization except for the states
near the conductance peaks.

After obtaining all  eigenstates through exact diagonalization,
we calculate  $\sigma_{xy}$  based on  Kubo formula:
\begin{eqnarray}
\sigma=\frac{i\hbar e^2}{A}\sum\limits_{m,n}
(f_{m}-f_{n})\frac{\langle m \vert v_x \vert n\rangle\langle n \vert v_y \vert m\rangle-h.c.}{(E_m - E_{n})^2 + \eta^2},
\label{kubo}
\end{eqnarray}
where $A$ is the finite-size system area,
$\mathbf{v}$ is the velocity operator, and $f_{m(n)}$ is the
Fermi-Dirac distribution function.  $\vert n\rangle$ 
and $\vert m \rangle$ denote exact eigenstates of the system.
$\eta=1/\tau_{in}$ is usually introduced to take into account of
the  finite relaxation time of the system due to inelastic scatterings.

{\bf Hall and longitudinal conductances at T=0---}We first present  zero-temperature results
by setting  $\eta=0$, where  $\sigma_{xy}$ is well
defined associated with the topological invariant Chern numbers\cite{huo,aro,lattice,dns1} of the states
and the  critical scaling behavior of the system is fully determined by disorder scattering.
As we tune the Fermi energy $E_f$, one can determine
a quantum phase transition by following the evolution of 
Hall conductance $\sigma_{xy}$, which is shown in Fig. \ref{hall1}(a) for $t_2=1/3$ and $W=2$.
We find that $\sigma_{xy}$  continuously  increases from
zero (insulating state) to the quantized value $e^2/h$, with data for different
system sizes approximately crossing each other at one single energy $E_c$.
The transition from the insulating state to Hall plateau state
becomes sharper with the increase of the system length $L$.

Now we consider  system size dependence of $\sigma_{xy}$ to
determine the scaling behavior of the quantum phase transition.
According to the scaling theory of the quantum Hall system,
 localization length should satisfy a powerlaw behavior\cite{review}
near a quantum phase transition as $\xi (E_f) \propto |E_f -E_c|^{-\nu}=|\Delta E|^{-\nu}$,
and the conductance should be a function of the single parameter
as the ratio $L/\xi(E)\sim L|\Delta E|^{\nu}$,
or equivalently  the scaling variable 
$(L/L_0)^{1/\nu} \Delta E$ (here $L_0$ is a constant for length scale) for large enough $L$.  As we 
replot data from different system sizes using the scaling variable as shown in Fig. \ref{hall1}(b), 
 all  data collapse onto one curve with $\nu=2.5$ giving best results,  consistent with the well accepted value for the insulator to plateau transition in a magnetic field.
Interestingly,  similar Hall conductance results are obtained
for the wider band with strong disorder ($t_2=0$ and $W=6$).
We plot all  data from both cases together in Fig. \ref{hall2}(a) as a function of the
scaling variable $(L/L_0)^{\nu}\Delta E$ with constant $L_0=24$ and
 $\nu=2.5$.
The transition in the stronger $W$ case is much slower, however all data points from both
systems can also be collapsed together through rescaling
the $x$ variable to  $b(L/L_0)^{1/\nu} \Delta E$.
We obtain $b=0.40$ for  the latter case (if we set $b=1$ for $W=2$ case), 
 indicating the length scale  in the latter is
relatively bigger as $L_0/b^{\nu}\sim 9.88L_0$. 
  Two curves merge into each other around the transition region,  consistent with the universal scaling
curves discussed before for the IQHE system under a magnetic field\cite{dns1}.
Furthermore, we find that near the transition region, the reflection symmetry
for Hall conductance as: $\sigma_{xy}(-\Delta E)=1-\sigma_{xy}(\Delta E)$ is 
satisfied despite
there is no particle-hole symmetry about the transition point ($E_c$) in the original Hamiltonian.

Now we discuss the scaling behavior for  Thouless  conductance $\sigma_T$\cite{thouless2}.
It has been established that the $\sigma_{T}$ is proportional to the longitudinal conductance\cite{mac1}
in the quantum Hall transition region and thus they satisfy the same scaling law. 
Shown in Fig. \ref{thouless1}(a),  we obtain $\sigma_{T}$ for different system sizes.
 The $E_c$ as the crossing point for $\sigma_{xy}$ of  different
sample sizes ($L=24$ to $L=120$)  also
coincides with the peak position  of  Thouless conductance $\sigma_{T}$.
  The $\sigma_{T}$ at the critical point shows a slow increasing
with system length,  which scales to a constant value at large $L$ limit.
Away from the peak position, $\sigma_T$ decreases with the increase of $L$.
This is consistent with a direct transition
with the localization length of the system  diverging  at $E_c$.
To check if the  $\sigma_{T}$ satisfies the one parameter scaling law,
we plot the renormalized conductances as a  function of the scaling variable $(L/L_0)^{1/\nu}\Delta E$
with the same exponent $\nu=2.5$.
We see clearly that all  data points are collapsing onto one curve, except  data from the smaller size $L=24$.
We also find that the symmetry relation $\sigma_{xx}(-\Delta E)=\sigma(\Delta E)$ is only satisfied in a small
region near $E_c$. More carefully examination suggests that the shape of the scaling curve is slowly changing with the increase
of $L$ to make the curve more symmetric about the transition point. Thus we believe that, the particle-hole
symmetry will be recovered over larger range of the transition at large $L$ limit.
We further obtain the $\sigma_{T}$ for the  wider band case with $t_2=0$ and $W=6$.
Our results are  more or less symmetric about the critical point $E_c$.
  The $\sigma_{T}$ is near constant at the critical point,
while all $\sigma_{T}$ values drop with the increase of $L$ away from $E_c$.
Furthermore, we replot all data points using the scaling variable $(L/L_0)^{\nu}\Delta E$
and indeed all  data seem to collapse onto one curve as long as $L>48$.
The deviation from  the scaling at smaller sizes is not surprising since the length scale
in this wider band  case is about one order of magnitude bigger than $t_2=1/3$ and $W=2$ case.
We also observe a stronger  trend of adjusting of the shape of the scaling curve
for larger system sizes toward more symmetric curve
(it drops faster at $\Delta E<0$ side and slower at the other side for $L=120$ system).

Through these simulations,  we establish that for the topological band model we studied with strong disorder, the insulator to plateau
transition indeed demonstrates  the same scaling law as the IQHE under magnetic field
 consistent with the  scaling theory\cite{pruisken, review} for the unitary class. It is
important to see that even at finite system sizes we can examine here ($L\sim 120$),
the one parameter scaling law is well established for both $\sigma_{xy}$ and $\sigma_{T}$,
while for the latter there is a slow tuning of the shape of the scaling curve towards more
symmetric one upon the increase of the system length.
We have also checked that for weak disorder limit,  there are much wider region of the energy band where states
appear to be delocalized or with localization length much longer than our system sizes. One needs to
go to much larger $L$ to see systematic scaling behavior,  which is beyond the scope of our current work.

{\bf Hall conductance with finite relaxation time and the finite temperature effect--}Now,  we are ready to  study the temperature dependence of the quantum critical behavior.
For the finite temperature transport,   one has to take into account the\cite{emil1, noncommute}
 finite $\eta=1/\tau_{in} \propto (kT)^p$ 
due to the finite relaxation time $\tau_{in}$ caused by the temperature dependent 
inelastic dephasing process.
Theoretically,  based on the scaling argument\cite{thouless2,review}, it is known that 
the inelastic scattering   exponent  $p=1$ for noninteracting  and $p=2$
for electron-electron interacting systems\cite{review}. For disorder systems we study,  it is not well established 
how the finite relaxation  time  will affect the Hall conductance, which has been studied recently based
on the noncommutative Kubo formula\cite{emil1}.
Here, we take a different approach by following  Kubo formula Eq. (1) as it is accurate for large system sizes
we consider, which also provides the advantage of studying length scaling and finite T scaling on equal footing.
As a start, we will  take  $1/\tau_{in}$ as a free parameter for Eq.(1) and calculate the
corresponding $\sigma_{xy}$ at $T=0$.
In Fig. \ref{eta}(a), we show the overall behavior
of  the  derivative $(d\sigma_{xy}/dE_f)|_{E_c}$ at the transition point as a function
 of the $1/\tau_{in}$,
which characterizes the sharpness of the insulator to plateau transition.
  As indicated by the straight line fitting in the logscale plot, we find that 
 $(d\sigma_{xy}/dE_f)|_{E_c} \propto (1/\tau_{in})^{x}$,  with the exponent $x=0.22 \pm 0.03 \simeq 1/(2\nu)$
for two systems $L=60$ and $L=120$ in a wide range of $1/\tau_{in}$.  The relatively larger error bar
is due to the sensibility of the derivative on small changes of the $\sigma_{xy}$ and we have taken
more than 1000 disorder configurations average to ensure accurate results.   At high $1/\tau_{in}$ side,
the scaling fails around $1/\tau_{in}> 0.20$, where the effective length
of the system $L_{in}$ is reduced to a couple of lattice constant determined by a 
comparison with the Hall conductance of small $L\sim 6$ size  with $1/\tau_{in}=0$.  On the other hand,
at  small $1/\tau_{in}$ limit,  the derivative saturates to a near constant
due to the fact that the effective length of the system is being cut-off by the
sample length when $L \leq  L_{in}$,  thus the Hall conductance becomes insensitive
to the decreasing of the $1/\tau_{in}$.   Indeed, we see that the small $1/\tau_{in}$ cut off 
 is moving towards lower value with the increase of the sample length $L$.

  According to the  one parameter scaling law,  one
would expect that  $d\sigma_{xy}(T)/dE=\frac {df((L_{in}/L_0)^{1/\nu}\Delta E)} {dE} \propto (L_{in}/L_0)^{1/\nu}f(0)$,
where $f(x)$ is the  scaling function for the Hall conductance.
Our results suggest that indeed the finite $1/\tau_{in}$ effect is setting a 
finite dephasing length $L_{in}\propto \sqrt{\tau_{in}}$ as long as the $L$ is large enough
 so that $L_{in} << L$.
The observed  powerlaw behavior provides an explanation  to the experimental results\cite{exp} of
the powerlaw scaling of  $(d\sigma_{xy}/dE_f)|_{E_c}\propto T^{-\kappa}$. 
Because $1/\tau_{in}\propto (kT)^p$,  one obtains $\kappa \simeq p/2\nu \simeq 0.4$ or $0.2$ if we take
$p=2$ due to electron-electron scattering or $p=1$ for noninteracting systems, respectively.  
However, there could be a finite temperature correction
due to the contribution from the Fermi-Dirac distribution function.  
To explicitly address this effect,  we calculate the finite temperature Kubo conductance with different
strength of the inelastic scattering  $1/\tau_{in}=ckT$  by choosing $p=1$ for our noninteracting system.
As shown in Fig. \ref{eta}(b),  we find that the $(d\sigma_{xy}/dE_f)|_{E_c}\propto T^{-\kappa}$, with fitting
$\kappa\simeq 0.225 \pm 0.03$ for our system $L=120$ independent of the strength of the 
inelastic scattering $c$.
Furthermore,   there is a visible higher temperature break down, which sets in earlier for 
smaller $c$
as the Fermi-Dirac distribution will contribute more significantly for such systems
due to the sharper transition (or stronger dependence on $E_f$) near the transition point.  
 The cut-off temperature at low $T$ limit  depends  on the parameter $c$ and
the stronger inelastic  scattering gives wider range of the powerlaw scaling since the dephasing
length $L_{in}$ reaches $L$ at lower T.

We have also obtained  $\sigma_{xy}$ at different $E_f$ for different temperatures
as shown in Fig. \ref{eta-T} for parameter $c=2$. 
We find that all  data from a wide range of temperature
with k$T$ varying from 0.0002 to 0.02 in units of hopping $t$,  can  be collapsed onto one
 curve using the scaling variable $(T/T_0)^{-\kappa} \Delta E$, with the best fitting $\kappa=0.22\sim 1/2\nu$ as
expected from the scaling behavior of the $(d\sigma_{xy}/dE_f)|_{E_c}$.
We suspect that the small difference between the obtained $\kappa$ and $1/2\nu=0.2$
is due to the finite size effect (since $L_{in}<< L=120$).

To summarize,  we have systematically studied  zero temperature and finite temperature scaling behavior of
the insulator to plateau transition in topological band model.  While we observe
universal scaling behavior for  zero temperature Hall and longitudinal conductances,
we also find that the wider band with stronger disorder has much larger length scale
for reaching the one parameter scaling regime.  At low enough temperature, the Hall conductance follows 
 one parameter scaling law:
$\sigma_{xy} =f((T/T_0)^{-\kappa} \Delta E)$, with $\kappa=p/2{\nu}$
fully determined by  temperature dependence of  inelastic relaxation time
$1/\tau_{in} \propto T^p$.  Our results suggest that the electron-electron interaction
is relevant for  experimentally observed $\kappa$  through
its temperature dependent relaxation time ($p=2$)\cite{exp} while
the length scaling exponent $\nu$ remains unchanged by such inelastic relaxation effect.

{\bf Acknowledgments} - We thank Emil Prodan for discussions.
This work is supported by US National Science Foundation  Grants 
DMR-0906816, PREM grant DMR-1205734, and Princeton MRSEC Grant DMR-0819860 for travel support.

\end{document}